\documentclass{aa}
\usepackage{epsfig}

\usepackage{graphicx}

\newcommand{\teff}{$T_{\rm eff}$}

\newcommand{\logg}{log\,$g$}
\newcommand{\kms}{km\,s$^{-1}$}

\newcommand{\ea}{et~al.~}

\newcommand{\msun}{$M_{\odot}$}
\newcommand{\rrls}{RR\,Lyrs}
\newcommand{\rrl}{RR\,Lyr}
\newcommand{\hipp}{{\sc Hipparcos}}

\begin{document}
   \title{RR Lyrae stars: Kinematics, orbits and $z$-distribution }
   \subtitle{}

   \author{G. Maintz
          \and K.S. de Boer
         } 

   \institute{Sternwarte der Universit\"at Bonn,
            Auf dem H\"ugel 71, 53121 Bonn, Germany\\
             \email{gmaintz,deboer@astro.uni-bonn.de}
           }

   \abstract{RR Lyrae stars in the Milky Way are good tracers 
to study the kinematic
behaviour and spatial distribution of older stellar populations. A recently
established well documented sample of 217 RR Lyr stars with $V<12.5$ mag,
for which accurate distances and radial velocities as well as proper motions
from the Hipparcos and Tycho-2 catalogues are available, has been used to
reinvestigate these structural parameters. The kinematic parameters allowed
to calculate the orbits of the stars. Nearly 1/3 of the stars of our sample
has orbits staying near the Milky Way plane. Of the 217 stars, 
163 have halo-like orbits fulfilling one of the following
criteria: $\Theta < 100$ \kms, orbit eccentricity $>0.4$, and normalized
maximum orbital $z$-distance $>0.45$. Of these stars roughly half have
retrograde orbits. The $z$-distance
probability distribution of this sample shows scale heights 
of $1.3\pm0.1$\,kpc for the
disk component and $4.6\pm0.3$\,kpc for the halo component. 
With our orbit statistics method we found a (vertical) spatial 
distribution which, out to $z=20$\, kpc, is similar
to that found with other methods. This distribution is also 
compatible with the ones found for blue (HBA and sdB) halo stars. The
circular velocity $\Theta$, the orbit eccentricity, orbit $z$-extent and
[Fe/H] are employed to look for possible correlations. If any, it is that
the metal poor stars with [Fe/H]\,$<1.0$ have a wide symmetric distribution
about $\Theta=0$, thus for this subsample on average a motion independent of
disk rotation. We conclude that the Milky Way possesses a halo component of
old and metal poor stars with a scale height of 4-5\,kpc having random
orbits.  The presence in our sample of a few metal poor stars 
(thus part of the halo population) with thin disk-like orbits 
is statistically not surprising. The
midplane density ratio of halo to disk stars is found to be 0.16, a value
very dependent on proper sample statistics.
}
   \maketitle

\section{Introduction}
\label{intro}

Studying the spatial distribution of stellar populations in the Milky Way 
requires well defined tracer objects. Among the large variety of stars,
those of horizontal-branch (HB) nature are well suited. 
HB stars are core-He burners after the red-giant phase whose envelopes 
have little ($<0.1$\,\msun) to perhaps 0.7\,\msun, 
leading to a location on the HB from the very blue to red 
(sdB, HBB, HBA, RR Lyr, RHB stars). 
They had an initial mass on the main sequence of $<2.5$\,\msun 
which means that some of the \rrls\ are only about 1 Gyr old and 
therefore belong to the younger disk population. 
In particular the blue HB stars and RR Lyrae variables are easy to find.
For the RR Lyrae stars (\rrls) the mean absolute magnitude
is well calibrated using data from the \hipp\,(de Boer \ea 1997a;
Fernley \ea 1998).  
The distances of blue HB stars  are easily determined utilising 
their  well defined  $M_{V}$. 
For the blue HB stars, models based on the evolutionary nature 
allow to derive \teff\,
and \logg\, from their spectra, thus leading to a fairly accurate $M_{V}$ 
as well. Red HB stars are less easy to identify, but the \hipp\, data 
allow to isolate nearby ones from the parallax based CMD (Kaempf \ea 1995).

As a result of many years of work on the nature and distribution
of sdB stars (Heber 1986; Moehler \ea 1990; Theissen \ea 1993; 
Saffer \ea 1994, Villeneuve \ea 1995) 
the idea emerged to include radial velocity and proper
motion data allowing to calculate the kinematical properties of these
stars, including the calculation of orbits 
(following Odenkirchen \& Brosche 1992). 
Several such studies have been carried out thus far.

Using sdB stars, de Boer \ea (1997b) derived from the orbit 
statistics of 41 stars a $z$-distribution with a scale height 
of $\simeq1$\,kpc plus a hint 
of the existence of a more extended population. 
The existence of this latter population was 
substantiated by Altmann \ea (2004) in a sample of 114 stars, 
showing a halo group with a $z$-distribution
which one can characterise by a scale height of  $\simeq7$\,kpc.
Recently, using the CMD identification of red HB stars,
Kaempf \ea (2005) presented a similar study of the RHB star distribution. 
They find a disk-like distribution with a $z$-scale height of perhaps 
up to 0.8\,kpc and an indication for an extended 
population with a scale height of $\simeq 5$~kpc.

It is therefore natural to carry out a similar study for 
RR Lyrae stars since their nature is intermediate to RHB and sdB stars. 
This has become reliably possible using the recent magnitude 
limited ($V<12.5$\,mag) complete catalogue of RR Lyrae stars (Maintz 2005),
a catalogue free of all kinds of misidentification.

Sect.\,\ref{data} describes the basic data and in Sect.\,\ref{kinorb}   
we present the results of our calculations. We analyse the 
velocities in Sect.\,\ref{anavel} as well as correlations between 
velocity and other parameters like metallicity. The implications 
of our findings are discussed in Sect.\,\ref{scaleres} 
and a comparison with results from other studies 
in Sect.\,\ref{otherstudies}.

\section{Data}
\label{data}
\subsection{The sample}
\label{sample}
The \rrls\ used for the present investigation are taken from the well
documented sample established by Maintz (2005). She checked the
identification of \rrls\ of the GSC with $V<12.5$\, mag against finding
charts and with new observations.

For the present investigation we restricted ourselves to those \rrls\ from
her list which were observed in the Hipparcos mission (see The Hipparcos and
Tycho Catalogues, ESA 1997; {\bf Hip} and the Tycho 2 catalogue, H{\o}g \ea
2000 {\bf Tyc2}). Our sample thus has the best positions and proper motions
available. Furthermore, radial velocities have to be known. Of the 104
\rrls\ found in the Tyc2 we used 62, the other 42 having no reliable $v_{\rm
rad}$. 
 From the Hip we could use 154 of the 182 \rrls\ for the same reason.
Our data for radial velocity  as well as distance from the galactic
center came from sources in the literature (Beers \ea 2000, Fernley \ea 1998,
Layden 1994).
We note that these authors used slightly different methods 
to come to stellar distances.
We thus have investigated 217 \rrls\ for their kinematic behaviour.
Our sample has no kinematical bias. 
The limits are the limiting magnitude of the Hip and Tyc2 catalogues, 
combined with the availability of the $v_{\rm rad}$ data.

\begin{figure}
  \begin{center}
  \epsfig{figure=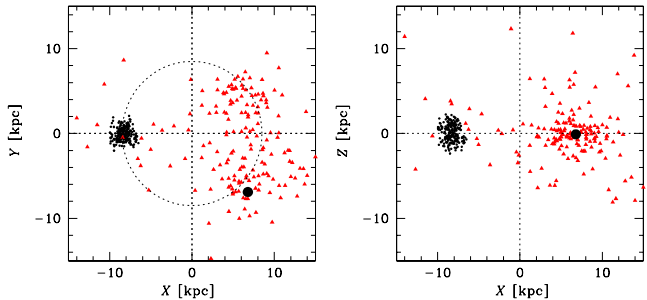,scale=0.45}
  \caption{Distribution of the \rrls\ of our sample today (hexagons)
 and 100 Myr ago (triangles), equivalent to about half a galactic revolution. 
 The full circle shows the position of the Sun 100 Myr ago. 
  At left the distribution projected on the Galactic plane,
  at right perpendicular to the plane (along the $X$-axis).
       The circle in the left panel shows the 8.5~kpc radius. 
 }
 \label{xyzpos3804.fig}
 \end{center}
\end{figure}

The present location of the  \rrls\ of our sample in the $X,Y$ and 
in the $X,Z$ plane is shown in Fig.\,\ref{xyzpos3804.fig} with hexagons. 
Evidently, \rrls\ have been searched for and found foremost 
at higher galactic latitudes, avoiding the crowded region of the disk. 
Furthermore, since our sample is magnitude limited, those stars included
while suffering from interstellar extinction have a smaller distance for
their apparent magnitude. 

The \rrls\, being located now in the solar neighbourhood spread with time
to nearly any galactic location 
(see Fig.\,\ref{xyzpos3804.fig}, triangles). 
Many must have been at almost any location, both at very small and very large 
galactocentric radius and at small and large $z$-distance. 
 This means, that our sample of 217 \rrls\ contains stars one can label
as part of a halo, of a thick disk, and of a thin disk population
(not claiming these are sharp population distinctions).
The last category has kinematics keeping them confined to the galactic plane
and thus they may be underrepresented in our sample.
Among the 217  are, however, several
\rrls\ with solar abundances (see Sect.\,\ref{Ferrl} and
Sect.\,\ref{sorbits}), suggesting that the lack of stars with thin disk
kinematics need not be excessive.

\subsection{Metallicity of the \rrls}
\label{Ferrl}

\begin{figure}
  \begin{center}
  \epsfig{figure=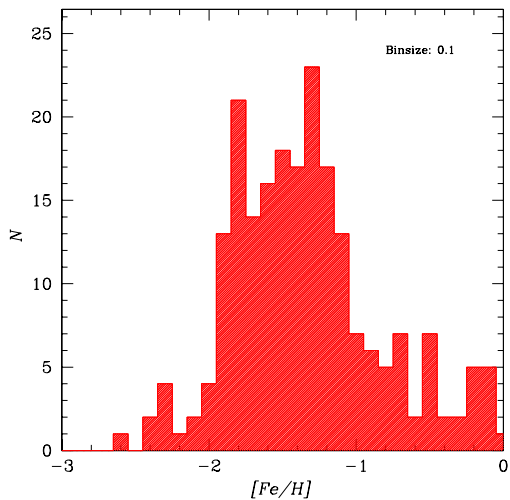,width=70mm}
  \caption{Histogram of the distribution of our stars in [Fe/H].
\rrls\ cover a range from solar like values to very metal poor.
 Most stars show metallicity $-2<$~[Fe/H]~$<-1$. Note that 
several stars have metallicities near the solar value.
}
 \label{Fehist.fig}
 \end{center}
\end{figure}

The metallicity of our stars is taken from the literature. 
We use data given by Beers \ea (2000), Fernley \ea (1998) and  Layden (1994).
As shown in Fig.\,\ref{Fehist.fig}, the metallicity of \rrls\ ranges from
solar values to very low values of $-$2.6~dex. 
A broad peak in Fig.\,\ref{Fehist.fig} shows that most of the stars 
of our sample have metallicities of $-2<[\rm Fe/H]<-1$. 
A small tail represents stars with still
lower metallicity. These abundances indicate, following the usual arguments,
that most of the stars are of population II.
But there are \rrls\ with higher metallicity as well: 46 stars
have  [Fe/H]~$>-1$.
We have in our sample 27  \rrls\ with a metallicity of
[Fe/H]~$>-0.7$ and 11 stars even have [Fe/H]~$>-0.3$.
This suggests that the lack of relatively young \rrls\ 
(likely having kinematics keeping them near the plane) cannot
be  excessive.

\section{Kinematics and orbits}
\label{kinorb}
We have analysed  the distribution 
of various kinematical aspects of our sample. 
To get the information for the overall kinematics of the \rrls\ 
and to attempt to come to a separation of disk and halo  we
calculated their orbits. 

\begin{figure*}

   \centering
   \epsfig{file=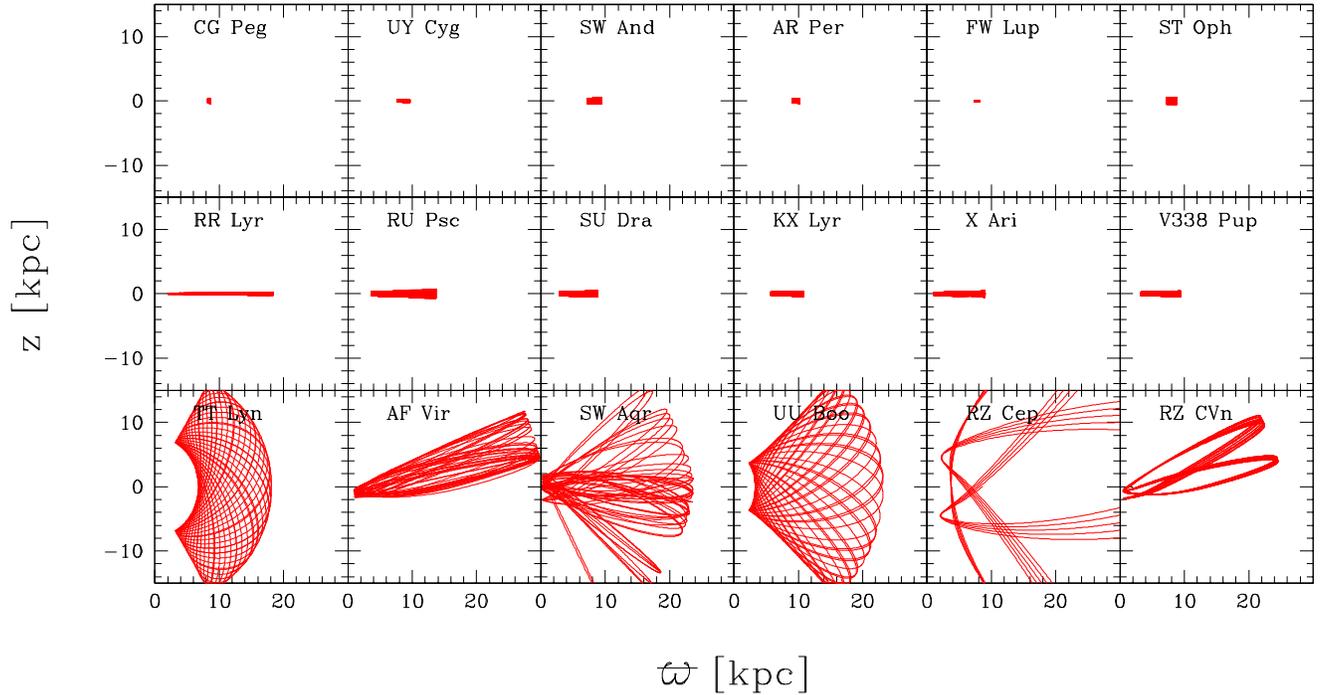,width=175mm}
   \caption{Meridional plots of some of the calculated RR Lyrae stars orbits. 
    The orbits are calculated over a times pan of 10~Gyr. 
    The top row shows  orbits of thin disk star candidates. 
    The  middle row shows the orbit of RR Lyrae  itself (on the left)
    and five more stars with elliptic orbits in/near the plane. 
    The bottom row shows orbits of stars 
    moving far out into the halo. 
      For the orbit of the Sun see de Boer \ea (1997b, their figure\,1).
        }
   \label{orbits.fig}
\end{figure*}

\subsection{Calculating velocities and orbits}
\label{kinorbcal}

The observational parameters 
$\alpha,\delta,d,\mu_{\alpha},\mu_{\delta},v_{\rm rad}$ were transformed into
$X,Y,Z,U,V,W$ (for details see de Boer \ea 1997b). 
Also the orbital velocity projected to the galactic plane, $\Theta$,
and the velocity towards the galactic centre, $\Phi$, were calculated. 
Orbits were calculated over a time span of 10~Gyr\footnote{This long time 
allows to show more clearly  the shape of the orbits. }
taking steps of 1\,Myr.
Following de Boer \ea (1997b) the eccentricity, $ecc$, 
of the orbit and the apo- and  perigalactic distances, 
$R_{\rm a}$ and $R_{\rm p}$, were calculated as well as the maximum 
$z$-distance reached, $z_{\rm max}$, 
and its galactic radial distance normalised $z$-extent, $nze$, 
\begin{equation}
ecc=\frac{R_{\rm a}-R_{\rm p}}{R_{\rm a}+R_{\rm p}} \ \ \ {\rm and} \ \ \ 
nze=\frac{z_{\rm max}}{\varpi(z_{\rm max})}
\end{equation}
For the values for our stars see Table \,\ref{tablecds}\, 
available in electronic form.

\subsection{The orbits and the orbit parameters}
\label{sorbits}

The  meridional sections of the orbits of \rrls\ (see Fig.\,\ref{orbits.fig}) 
show very different shapes.
 About half of the stars of our sample have perigalactic distances $R_{\rm p}
\leq 3$\,kpc.
  65 \rrls\ reach a perigalactic distance $R_{\rm p}\leq 1$\,kpc. Three stars
(DR\,And, AM\,Vir and AV\,Vir) reach perigalactic distances less than 0.1~kpc.
 The other extreme is represented by 5 \rrls\ (CI\,And, AL\,CMi, TW\,Lyn,
AR\,Per and SS\,Psc) 
which have  perigalactic distances more than 9 kpc and 
always stay   beyond the solar circle.

Of the 217 \rrls\ 63 have boxy orbits and stay close to the plane.
A subset of 29 makes only small excursions in $\varpi$ 
(Fig.\ \ref{orbits.fig} top row), 34 have planar but very eccentric 
orbits (Fig.\  \ref{orbits.fig} middle row).  
The orbit of \rrl\, itself is not like that of a disk population star 
(even although its $z_{\rm max}$ is only 0.21\,kpc and its $nze = 0.01$), 
because its orbit has $ecc =0.8$ with $R_{\rm a}=18.4$\,kpc 
and $R_{\rm p}= 2.1$\,kpc.

154 of the  \rrls\ show orbits resembling those of the bottom row.
The shape of these  orbits is really chaotic, and shows 
movements  perpendicular to the plane. 
These stars partly have orbits going to very small galactocentric
distances and some of them  reach very large apogalactic distances,
 with as an extreme $R_{\rm p} \simeq 51$\,kpc (for SX\,Aqr).

\begin{figure}

  \begin{center}
  \epsfig{figure=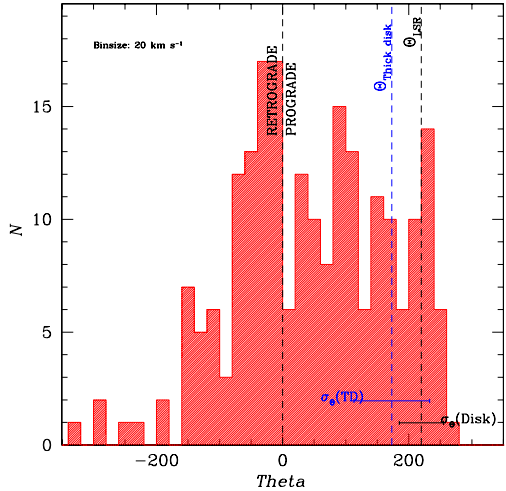,width=70mm}
  \caption{Histogram of the distribution of our stars in ${\Theta}$
($\Theta_{\odot}=220$ \kms). Some stars show circular velocities 
regarded to be characteristic for the thick and thin disk, respectively 
(see dashed lines). 
${\Theta}_{\rm LSR}$, ${\Theta}_{\rm thick\ disk}$
and the velocity dispersions for thick and thin disk 
as found by Ojha \ea (1994) are indicated. 
Most \rrls\ are halo stars and the high peak represents the \rrls\ 
 with retrograde motion. 
 }
 \label{thetahist.fig}
 \end{center}
\end{figure}

\begin{figure}

   \centering
   \epsfig{file=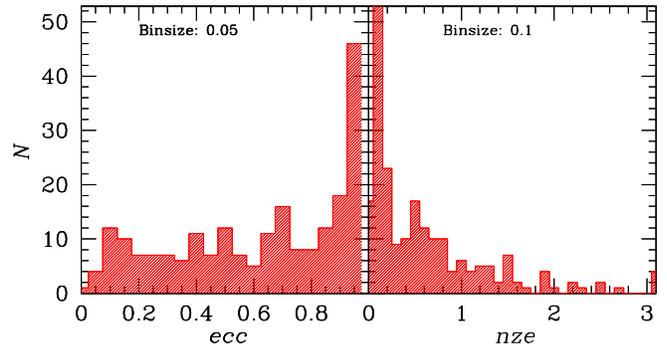,scale=0.45}
  \caption{Histograms showing the distributions of $ecc$ (at left) and $nze$
           (at right) for the RR Lyrae stars of the sample. 
            Note the flat distribution of  $ecc$ at low values
          and the peak  at  $ecc \simeq 0.9$.
           Values of $nze$  lie predominantly between 0 and 1,
         with a local minimum at $nze\simeq0.4$ and a long tail up to $nze=3$.
        }
  \label{eccnzehist.fig}
\end{figure}

The statistics of the circular component $\Theta$ of the velocity is 
shown in Fig.\,\ref{thetahist.fig}. 
In our sample we have 87 \rrls\ with retrograde orbits.
We compared the characteristics of these stars with the prograde ones.
Among the retrograde part of our sample there are no stars with a metallicity
[Fe/H]~$>-0.9$ (see Sect.\,\ref{anavel} upper left panel).
Considering the retrograde group separately shows that the peak at high
eccentricities in Fig.\,\ref{eccnzehist.fig} (left panel) is nearly
completely due to the retrograde  \rrls, while the stars with prograde
orbits show a flat distribution over the hole metallicity range.

The eccentricities $ecc$ of the orbits as well as the values of $nze$ 
span a large range (see Fig.\,\ref{eccnzehist.fig}).

The eccentricity of the orbit of the majority of our stars is $ecc~>0.45$. 
The distribution has an absolute maximum at $ecc= 0.9$ and a
shallow local minimum at $ecc\simeq 0.6$.

The distribution of the normalised $z$-extents shows a maximum at low values
$nze<0.2$, a local minimum at $nze =0.4$ followed by another peak at
$nze \simeq 0.6$. There are 109  \rrls\ with $nze \leq 0.4$ and only 45 \rrls\
reach $nze \geq 1$. A separation in stars with prograde and
retrograde orbits does not change the appearance of the $nze$ diagram. 
Both groups show high peaks at low values of $nze$ and a long flat tail.

\subsection{Kinematic parameters and population type}

The shape of a stellar orbit gives an indication of the population a star
belongs to. 
{\sl Disk-population} stars have orbits similar to those of the younger 
galactic stars, being rather circular and staying close to the plane. 
These are the ones most likely having been born in the disk. 
It is, from our data, not easy to discriminate between so-called 
thin- and thick-disk stars, a distinction which is not well defined anyway. 
The {\sl halo population} consists of stars whose orbits are not 
akin to disk-like galactic rotation. 
They most likely have been born outside the disk of the Milky Way. 

Using the above criteria (which were largely set by the distribution of the 
parameters of our stars as discussed above using  
Figs.\,\ref{thetahist.fig} and \ref{eccnzehist.fig}), 
one can now roughly sort the stars according to populations. 
We define halo stars as those having 
$\Theta<100$\,\kms\ or $ecc > 0.4$ or $nze>0.4$. 
Each one of the criteria alone suffices to make the star a non-disk born star, 
and thus a star belonging to the halo population. 

Our statistics shows that out of the total of 217 stars, 
140  stars have $\Theta < 100$\,\kms,
153 stars have $ecc> 0.4$, and 
108 stars have  $nze>0.45$,
while 163 stars fulfil at least one of the criteria.
Or, there are (kinematically) just 54 classic    disk 
stars in our sample.

\subsection{$z$-probability plot and scale height}
\label{scaleintro}

\begin{figure} 
  \begin{center} 
  \epsfig{figure=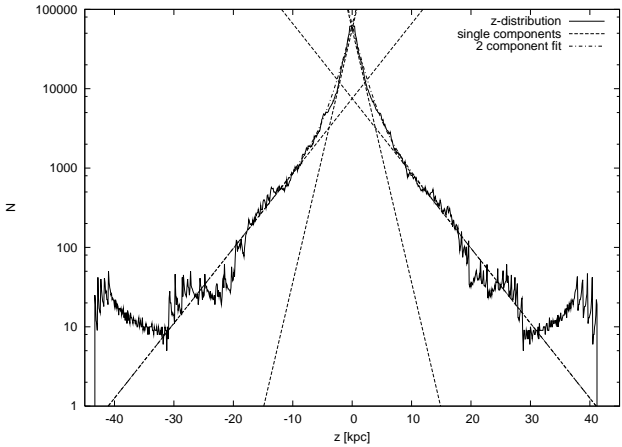,angle=270,width=83mm}
  \caption{The probability of the $z$-distance distribution  of
     all  stars of the sample,  obtained by summing the individual $N(z)$
      statistics for time steps of 1\,Myr.
      The dashed lines show the two-component  
       fit of the logarithmic distribution based on  the scale heights
      $h = 1.3 \pm 0.1$\,kpc for the ``disk component'' and
      $h = 4.5 \pm 0.3$\,kpc for the ``halo component''.
 } 
 \label{allergbest.fig} 
 \end{center} 
\end{figure}

The sum of the orbits was used to derive the associated 
probability distribution of the $z$-distances of our \rrl\ star sample.
This distribution is equivalent to the statistical $z$-density gradient 
of the sample (for further details see de Boer \ea 1997b).
One may fit the distribution with an exponential 
\begin{equation}
       \ln{N(z)}=\ln{N_0}-\frac{z}{h},
\label{barform}
\end{equation}
with $N_0$ being the density in the local galactic plane and $h$ the scale
height.
The slope of the $\log N$ vs $z$ distribution gives the scale height. 
Fig.\,\ref{allergbest.fig} shows our $z$-distance statistics. 
It turns out the left and right side are not identical 
(which is statistically not to be expected with the sample size we have) 
but the differences are small. 
We fitted the left and right side separately. 
The left side gives scale heights of $1.25\pm0.1$\,kpc for
the steep central part and $4.47\pm0.3$\,kpc for the flanks.
On the right side the values found are $1.30\pm0.1$\,kpc and 
$4.60~\pm0.3$\,kpc, respectively. 
We arrived at these values after several tries,
e.g., by changing the bin size and, because of small number statistics there, 
the limits of $z$ to be included. 
For the fit of the central steep part we used a fitting interval
of [0.3, 2.9]  and  for the flanks [5, 33].
Given the similarity of the results we finally give one overall fit 
with $1.3\pm0.1$\,kpc for the tight distribution 
and $4.5\pm0.3$\,kpc for the extended distribution. 
The ratio of the  mid-plane densities of the distributions is
 $N(0)_{\rm{halo}}$/$N(0)_{\rm{disk}}=0.16$.

The component with a steep $z$-distribution can be associated 
with the disk population, both the thin and thick disk. 
We tried a 3  
component fit to separate thin and thick disk but
did not arrive at sensible results. 
It is, from this data, not possible to separate thin and thick disk.

The analysis of the $z$-distribution statistics indicates
the presence of (perhaps at least) two populations.
The disk population consists of stars which stay near the plane,
while the halo population ventures to larger $z$. 
In order to explore a possible sample bias in relation with the scale height
of the halo population we have made two tests with subsamples, as follows.

First, we selected those stars which {\sl at present} are within
the disk, i.e., those stars with $z_{\rm present}<0.5$\,kpc.
We made the $z$-statistics of this sample, as well as of the complement,
and compared the two.
The large scale $z$-distribution of the $z_{\rm present}~<0.5$\,kpc group
shows a pronounced peak near $z=0$~kpc (which is of course what one expects).
The $z$-distribution of the $z_{\rm present}>0.5$\,kpc group resembles 
the distribution of the whole sample but the  mid-plane peak is missing.

The flanks of the $z$-statistics toward large $z$ of the two subsamples
are very similar, indicating there is no sample selection in this sense,
although the spatial distribution of the full sample is quite asymmetric
(see Fig.\,\ref{xyzpos3804.fig}).

 In the subsample of the $z_{\rm present}<0.5$\,kpc group we find a slight
separation of the thin and thick disk component leading to a scale height of
$0.38\pm0.03$\,kpc for the thin disk.
This value is only slightly higher than that 
given by Chen \ea (2001). 
We note that our $z_{\rm present}<0.5$\,kpc sample contains 63 stars.
Still, 
the separation of thin and thick disk is not well defined in our sample. 

Second, we made a subdivision based on $z_{\rm max}<0.5$ kpc.
Here the small $z$ subsample is by definition restricted to stars
staying close to the plane,
while the $z_{\rm max}>0.5$\,kpc subsample must contain 
predominantly halo stars.
The small $z$ subsample (see Tab.\,\ref{05zmax.tab}) shows 
in a $\ln N$ versus $z$ diagram only a small high peak near $z=0$\,kpc 
(by definition). This is further discussed in Sect.\,\ref{scaleres}. 

The large $z$ subsample shows a  $z$-statistics quite similar 
to that of the full sample.
It exhibits the  scale height of $4.6\pm0.3$\,kpc for the halo component, 
depending  on the choice of bin size and the limits of $z$ to be included.
That value is hardly different from the halo scale height obtained from
the full sample.
Even the disk component is present with the same values of scale height
as found for our full sample of \rrls, 
which is not surprising finding a
scale height of the disk component of 1.3\,kpc.
The presence of a disk component in this sample shows that 
the disk population is dominated by  stars with orbits reaching beyond 
 $z= 0.5$ kpc.

We thus conclude that there is no significant sample bias in our
\rrl\ star sample.

Altmann \ea (2004), dealing with sdB stars found with 
the same method a scale height of 0.9~kpc for the disk 
component and a halo component with a scale
height of 7~kpc and Kaempf \ea (2005), considering RHB stars, 
give a scale height of 5 kpc for the halo component.

\section{Sample statistics of $\Theta$ with ecc, nze, [Fe/H]}
\label{anavel}

\subsection{Circular velocity $\Theta$} 

The \rrls\ of our sample have a  mean circular velocity of 
$\Theta = 47$\,\kms, but the stars cover a wide range 
(see also Table\,\ref{RRvel.tab}) with  +277 and $-321$\,\kms\ 
as extreme velocities.

Looking at the histogram of $\Theta$ (Fig.\,\ref{thetahist.fig}) 
the most pronounced peak in the distribution
is the one near 0 \kms, marking the group which in a broad sense does not 
participate in galactic rotation. 

\begin{figure}
  \begin{center}
  \epsfig{figure=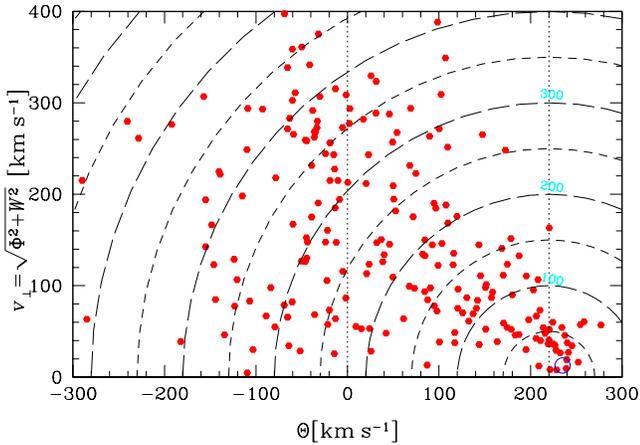,scale=0.43}
  \caption{Toomre diagram of ${\Theta}$ versus $v_{\rm perp}$, the velocity
perpendicular to the Galactic plane, showing the asymmetry in kinematics of
the \rrls. The dashed circles denote the peculiar velocity 
$v_{\rm pec}=\sqrt{\Phi^2+W^2+(\Theta-\Theta_{\rm LSR})^2}$ in \kms. 
The open circle shows the parameters of the Sun.
 }
 \label{toomneu.fig}
 \end{center}
\end{figure}

\begin{figure}
\begin{center}
   \epsfig{file=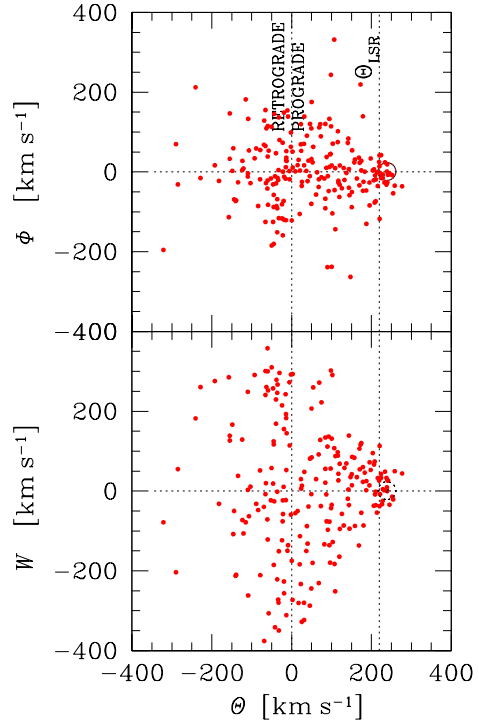,scale=0.50}
   \caption{Bottlinger and $\Theta-W$ diagrams  show the orbital
velocity of the RR Lyrae stars plotted against
the velocity towards the Galactic centre ($\Phi$, top) and
perpendicular to the Galactic plane ($W$, bottom).
A  circle indicates the Solar values.
 }
   \label{bottlinger.fig}
\end{center}
\end{figure}

\begin{table*}
\begin{center}
\setlength{\tabcolsep}{1.5mm}
\caption{Mean values of $U,V,W,\Theta, \Phi$, angular momentum, eccentricity 
and $nze$ with dispersions for (subsets of) the sample.}
\label{RRvel.tab}
\begin{tabular}{p{2.5cm}rrrrrrrrrrrrrrrrrr}
\hline
\\[-10pt]
  Subsample & $N$ &  $\bar{U} $ & $\sigma_U$ & $\bar{V}$ & $\sigma_V$ & $\bar{W}$ & $\sigma_W$
& $\bar{\Theta}$ & $\sigma_\Theta$ & $\bar{\Phi}$ & $\sigma_{\Phi}$ & $\bar{I_{\rm z}}$ &
$\sigma_{I_{\rm z}}$ & $\overline{ecc}$ & $\sigma_{ecc}$ & $\overline{nze}$ & $\sigma_{nze}$ \\
  &    & \multicolumn{10}{c}{[km\,s$^{-1}$]}& \multicolumn{2}{c}{[kpc\,km\,s$^{-1}$]} & & & &\\
\hline
all                   &  217  & $-$3  & 158 & +47  & 125 & +3    & 85 & +47   & 123 & +5  & 159 & +395 & 1027 & 0.61 & 0.30 & 0.63 & 0.79 \\
$[\rm Fe/H]~>-0.5 $   &  20   &  $-$2 &  48 & +206 & 31 & $-$3   & 38 & +204  & 32  & +3 &  53 &+1718 & 356 & 0.19 & 0.13 & 0.09 & 0.11 \\
$[\rm Fe/H]~>-1  $    &   46  &  0    &  82 & +165 & 93  &$-$3   & 53 & +168 &  90 & +4  &  82 &+1436 &  810 & 0.32 & 0.28 & 0.32 & 0.67 \\
$[\rm Fe/H]~\leq-1$   &  169  & $-$4  &173  & +15  & 113 & +5    & 92 & +14  & 110 & +6  & 175 & +113 &  890 & 0.69 & 0.25 & 0.71 & 0.79 \\
$[\rm Fe/H]~\leq-1.5$ & 86 & $-$25 & 177  & $-$13  & 114 & +5    & 97 & $-$13 & 108 & +27 & 181 & $-$105 &  880 & 0.72 & 0.23 & 0.67 & 0.61 \\ 
$ecc~\leq0.45^{a}$    &   58  & $-$15 & 54  & +195 & 40  &$-$19  & 64 & +196 &  40 & +19 &  51 &+1644 &  395 & 0.22 & 0.12 & 0.19 & 0.35 \\
$ecc~>0.45$           &  149  & +1    &186  & $-$2 & 92  & +11   & 91 & $-$3 & 88  & 0   & 188 &$-$20 &  722 & 0.79 & 0.17 & 0.81 & 0.85 \\
$P~<0.35$\,d          &   20  & $-$37 &133  & +44  & 116 & +33   &100 & +44  & 115 & +39 & 134 & +387 &  990 & 0.59 & 0.30 & 0.64 & 0.49 \\
0.35$<~P~<$0.55\,d    &  103  & +2    &164  & +68  & 137 & +2    & 76 & +68 & 135  & +1  & 165 & +583 & 1131 & 0.57 & 0.32 & 0.53 & 0.75 \\
$P~>0.55$\,d          &   91  & $-$1  & 157 & +22  & 108 & $-$1  & 92 & +21  & 106 & +3  & 159 & +170 &  861 & 0.68 & 0.26 & 0.74 & 0.87 \\
  \hline
\end{tabular} 
\end{center}\vspace*{-2mm}
a) without seven stars having retrograde orbits
\end{table*}

The Toomre diagram (Fig.\,\ref{toomneu.fig}) allows to see the high asymmetry
of our sample including many stars with peculiar velocities 
$v_{\rm pec}>100$\,\kms. 
This asymmetry is the so called asymmetric drift 
(for more on that see, e.g., the review by Majewski 1993).
\rrls\  with ${\Theta} \leq \Theta_{\rm LSR}$ dominate our sample 
and the dispersion of data points is large. 
Only few stars with ${\Theta}\geq{\Theta}_{\rm LSR}$ are present.
This is clearly correlated with the high eccentricity of the orbits of most
of the \rrls\ as discussed in Sect.\,\ref{sorbits}.

The Bottlinger diagrams (see Fig.\,\ref{bottlinger.fig}) show the 
asymmetry of data points as well. 
At low values of  ${\Phi}$ and  $W$ 
the \rrls\ do not concentrate near ${\Theta}_{\rm LSR}$ 
but have a great variety of kinematics. 
The  kinematic dispersion of $W$ exceeds that of ${\Phi}$.
The distribution of data points appears to be similar in the two panels. 
It seems to be symmetric in respect to both the ${\Phi}=0$ axis (upper panel) 
and the $W= 0$ axis (lower panel).

\subsection{Discussion of sample metallicities in relation with 
orbits and periods}
\label{metorbper}

Given the orbit parameters, it is worthwhile to relate them with 
other parameters of \rrls. 
For that, we have divided our sample in subgroups in various ways 
(Table\,\ref{RRvel.tab}). 
We used non kinematic selection criteria as metallicity and period
but the kinematic criterion eccentricity as well. 
 We choose as cuts $[{\rm Fe/H}]>-0.5$, 
$>-1$, $\leq-1$, and $\leq-1.5$,  
for $ecc \leq0.45$ and $>0.45$, 
and for period $P<0.35 $\,d,  $0.35<P<0.55$\,d and $P>0.55$ d.

{\bf Metallicity}
is not a kinematic parameter.  
Yet, metallicity gives no clear division between disk and halo stars. 
Stars with  $[{\rm Fe/H}] \geq -1$ may be members of the disk group 
in view of their $\bar{\Theta}$ (in particular those with 
 $[{\rm Fe/H}] \geq -0.5$). 
But the velocity dispersion $\sigma_\Theta \simeq 90$\,\kms\ is larger 
than expected for a pure disk population.
So some of the \rrls\ with  [Fe/H]~$\geq -1$ can be regarded to be halo stars.

 The more metal-poor \rrls\ with [Fe/H]$\leq-1$ are candidates for halo stars.
 This is  corroborated by the result of  
$\bar{\Theta}=14$\,\kms, $\sigma_\Theta=110$\,\kms,  
$\overline{ecc} =0.69$  and  $\overline{nze} =0.71$ for this group.
The velocity values of $\bar{U}$, $\bar{V}$ and  $\bar{U}$ for the
metal poor part of our sample are similar to several published halo samples
(see Martin \& Morrison (1998), their Table 5) even if 
in their works the cuts for halo stars are set at more metal poor levels.
Therefore we give the results for \rrls\ with  [Fe/H]$ \leq-1.5$, too.
For this group our results shows to be in good 
agreement  with  results of previous investigations 
(see Martin \& Morrison 1998, Chen 1999). 
This metal-poor group of \rrls\ has  $\bar{\Theta}=-13$\,\kms,  
i.e., a mean retrograde rotation with respect to the LSR\footnote{Note 
that results for $\Theta$ as given in the literature are sometimes based on 
$\Theta_{\odot} = 0$, sometimes on $\Theta_{\odot} =220$ \kms}.

{\bf Orbit eccentricity}
supposedly divides between younger and older stars 
(more gravitational interactions of disk stars lead to 
larger deviations from the originally circular orbits). 
The subdivision at $ecc=0.45$ shows the expected correlations with 
kinematic parameters (see Table\,\ref{RRvel.tab}; notably $\bar{I_{\rm z}}$). 
This subdivision does not correlate in a pronounced way with metallicity. 

Note that we 
did not include 7 stars with retrograde orbits in the subgroup 
with $ecc\leq 0.45$ (Table\,\ref{RRvel.tab}). 
These 7 have a 
mean circular velocity
 $\bar{\Theta}=-165$\,\kms\ with dispersion of $\sigma_\Theta =56$\,\kms.
Also Chiba \& Yoshii (1997) report the presence of halo stars 
with low eccentricity orbits. 
Both results show that low eccentricity does not always mean that a star is a 
disk member. The 
58 \rrls\ are good candidates for disk stars (thin and thick disk).

\begin{table}
\begin{center}
\setlength{\tabcolsep}{1.5mm}
\caption{ [Fe/H] and Period
with their dispersions for the 217 RR Lyrae  stars of the sample.}
\label{FEPER.tab}
\begin{tabular}{p{2.5cm}rrrrrr}
\hline
\\[-8pt]
Subsample             &  $N$    & $\overline{\rm [ Fe/H]}$ & $\sigma_{\rm [Fe/H]}$&  $\overline{\rm Period}$& $\sigma_{\rm Period}$ \\
                      &         &                &                  & \multicolumn{1}{c}{d}  \\
\hline
all                      & 217     &       $-$1.33  & 0.51 & 0.51 & 0.12 \\
$[\rm Fe/H]~>-0.5 $      & 20      &       $-$0.28  & 0.16 & 0.44 & 0.06 \\
$[\rm Fe/H]~>-1$         & 46      &       $-$0.55  & 0.30 & 0.44 & 0.09 \\
$[\rm Fe/H]~\leq-1$      & 169     &       $-$1.54  & 0.32 & 0.53 & 0.11 \\
$[\rm Fe/H]~\leq-1.5$    & 86      &       $-$1.80  & 0.23 & 0.56 & 0.12 \\ 
$ecc~\leq0.45^{a}$       & 58      &       $-$0.90  & 0.57 & 0.47 & 0.09 \\
$ecc~>0.45$              & 149     &       $-$1.48  & 0.36 & 0.53 & 0.11 \\
$P~<0.35$\,d             & 20      &       $-$1.39  & 0.48 & 0.29 & 0.07 \\
0.35$<~P~<$0.55\,d       & 103     &       $-$1.11  & 0.51 & 0.47 & 0.05 \\
$P~>0.55$\,d             & 91      &       $-$1.58  & 0.40 & 0.62 & 0.06 \\
  \hline
\end{tabular} 
\end{center}\vspace*{-2mm}
a) not including seven stars having retrograde orbits 
\end{table}

\begin{figure}
   \centering
   \epsfig{file=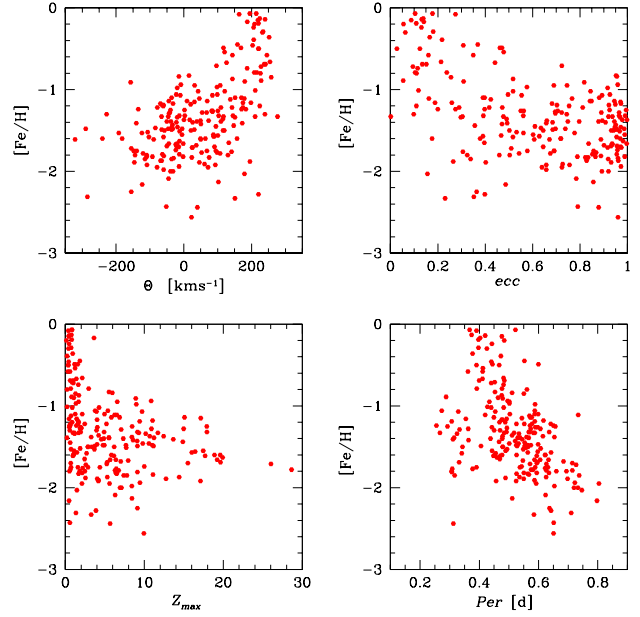,scale=0.44}
   \caption{The plot shows metallicity [Fe/H] versus circular velocity 
 $\Theta$  (upper left panel), $ecc$ (upper right), 
$nze$  (lower left) and period  (lower right panel). 
\rrls\ with [Fe/H]~$>-1 $  show more disk-like orbits,
 but more metal-poor stars present both, disk-like orbits and 
 halo-like orbits having smaller $\Theta$, large eccentricity $ecc$ 
 and large normalised $z$-distance $nze$.
 With [Fe/H] versus period (lower right panel), note that
there are 4 groups of stars. For the discussion see Sect.\,\ref{metorbper}.
}
   \label{FE4.fig}
\end{figure}

{\bf Pulsation periods}
are related with the mass of the \rrl, 
which is likely related with the development history of the star. 

The groups with pulsation period $P<0.35$\,d and $P>0.55$\,d show both 
lower abundances than the large third group of periods $0.35<P<0.55$\,d.
It has the largest abundance and the largest dispersion $\sigma_{\rm [Fe/H]}$.
It is not a uniform subsample of \rrls.
That this group is so different is also visible in Tab.\,\ref{RRvel.tab}. 
 The dispersion in $U$, $V$, $\Theta$, $\Phi$, $I_{\rm z}$ and $ecc$
is larger in this subsample than in the two other period groups.
This fact can be noted as well in Fig.\,\ref{FE4.fig} (lower right panel).

The group with periods of~$<0.35$\,d
contains the  RRc stars with abundances of~$-2<{\rm[Fe/H]}<-1$. 
The metallicity of the second period group ranges from 
solar to [Fe/H] of $\simeq-2$.
These are  RRab stars. Note that the period
increases in this group toward decreasing metallicity.
Most of the \rrls\ of our sample are in this group.
It combines \rrls\ 
with a large dispersion in metallicity 
and various orbit parameters, as can be recognised in Tab.\,\ref{FEPER.tab}, 
probably representing different ages.
The third group having periods $\simeq0.6$\,d are RRab stars as well but
with lower abundances. This group is the 
field \rrl\ equivalent to the \rrls\ in  Oosterhoff II clusters.
This last group is a very small subsample of \rrls\ with relative
long periods and metallicities [Fe/H]~$<-1.6$. 

{\bf All parameters}
are considered in Table\,\ref{FEPER.tab}. 
\rrls\, with  $ecc>0.45$ have lower abundances
($\overline {[\rm Fe/H]}=-1.48$) than those with $ecc<0.45$ 
($\overline{[\rm Fe/H]}=-0.90$),  as we  expected.
Like in Tab.\,\ref{RRvel.tab} we did not include 7 \rrls\ 
with retrograde orbits
having $ecc<0.45$. We did so because these stars do not fit into
the group of low eccentricity stars which are candidate disk stars.
This group of retrograde \rrls\ has a low metallicity 
of $\overline{[\rm Fe/H]} =-1.85$.

Fig.\,\ref{FE4.fig} shows the data of 
Tables~\,\ref{RRvel.tab} and \ref{FEPER.tab} in graphical form:  
metallicity [Fe/H] versus orbital velocity $\Theta$, $ecc$, $nze$, 
and period.
Dividing the \rrls\ in metal-poor and metal-rich  
groups  as Layden (1995), 
we also find that velocities for \rrls\ with [Fe/H]~$>-1$ reveal a 
distribution different than those with lower metallicities.
High metallicity stars with more disk-like circular velocities $\Theta$ 
cluster both at low $ecc$ and  $nze$. 
But the separation is not a strict one.
Even some data points with very low abundance [Fe/H]~$\leq-1.5$
show   $\Theta\simeq200$~\kms, $ecc<0.4$ and  $nze<0.2$.  
Altmann \& de Boer (2000) found the same fact using a much smaller 
sample of \rrls. 

These findings are similar to those by Chiba \& Yoshii (1998) 
who report that even stars with very low metallicity may have 
``nearly circular orbits with $ecc<0.4$\,''.
  High metallicity does indicate disk-like kinematics, 
but low metallicity does all by itself not give sufficient information 
about the orbit shape.
An eccentricity of $<0.4$ is alone not a sufficient criterion for population.
Chiba \& Beers (2000) find in their investigation of metal-poor stars 
``a remarkable discontinuity of the rotational properties of the Galaxy'' 
at  [Fe/H]~$\simeq-1.7$ (see also their figure 3). 
Our sample of \rrls\ does 
not   confirm this. 
We find a continuous distribution in
$\Theta$,  $ecc$ and  $nze$ in the metallicity range  $-2<[\rm Fe/H]<-1$.
But we note that in a sample of 163 metal-poor halo stars chances are
non-zero to find one with an orbit rather co-planar with the Milky Way disk.

\section{Discussion of results}
\label{scaleres}

We calculated the orbits of the \rrls\ of our sample and derived the 
$z$-distribution statistics. 
This led to scale heights for a disk group of \rrls\ 
($h_{\rm disk}\simeq 1.3$\,kpc) 
and for a halo group ($h_{\rm halo} \simeq 4.5$\,kpc). 
However, analysing for each star the parameters 
$\Theta$, $ecc$, $nze$, and [Fe/H] 
it became clear that all parameters are needed to classify a given star as 
a genuine disk or halo star. 
Of our 217 stars, 63 are truly disk like, 154 are halo like.

In our sample we have 87 \rrls\ with retrograde orbits.
We compared the characteristics of these stars with the prograde ones.
Among the retrograde part of our sample there are no stars with metallicity
[Fe/H]~$>-0.9$ (see Fig.\,\ref{FE4.fig} upper left panel).
These retrograde rotating \rrls\ are not separated from the rest of the sample
in respect to $z_{\rm max}$ or period, but their group values of  
$\overline{ecc}=0.79$, $\sigma_{ecc}=0.20$, 
$\overline{nze}=0.79$ and $\sigma_{nze}=0.79$ are 
higher than the values for prograde halo candidates of about 
the same metallicity. 
We again conclude that most of the \rrls\ of our sample are halo stars.

\begin{table}
\begin{center}
\setlength{\tabcolsep}{1.5mm}
\caption{16 \rrls\ staying close to the plane ($z_{\rm max}$$<$$0.5$\,kpc)
}
\label{05zmax.tab}
\begin{tabular}{lrrrrrr}
\\[-16pt]
\hline
\\[-8pt]
Name  &  $[\rm Fe/H]$ & $\Theta$ &$R_{a}$ & $R_{p}$ & $z_{\rm max}$ &$ecc$ \\
         &            &   [\kms]&  [kpc]  &  [kpc]  & [kpc]    &       \\
\hline
SW And   & $-$0.24 & 206.0    & 9.43  & 7.25 & 0.47 & 0.13 \\
AA CMi   & $-$0.15 & 219.1    & 10.45 & 8.77 & 0.43 & 0.09 \\
UY Cyg   & $-$0.80 & 226.6    & 9.61  & 7.70 & 0.23 & 0.11 \\
DM Cyg   & $-$0.14 & 240.1    & 10.59 & 8.11 & 0.48 & 0.13 \\
DX Del   & $-$0.39 & 206.4    & 9.84  & 6.31 & 0.23 & 0.22 \\
FW Lup   & $-$0.20 & 209.2    & 8.24  & 7.38 & 0.13 & 0.06 \\
RR Lyr   & $-$1.39 & 109.4    & 18.37 & 2.08 & 0.21 & 0.80 \\
CN Lyr   & $-$0.58 & 245.3    & 10.64 & 7.70 & 0.32 & 0.16 \\
KX Lyr   & $-$0.46 & 203.2    & 10.86 & 5.76 & 0.41 & 0.31 \\
AV Peg   & $-$0.08 & 163.6    & 8.63  & 4.91 & 0.39 & 0.27 \\
CG Peg   & $-$0.50 & 221.4    & 8.67  & 8.21 & 0.39 & 0.03 \\
AR Per   & $-$0.30 & 232.4    & 10.26 & 9.06 & 0.40 & 0.06 \\
HH Pup   & $-$0.50 & 220.2    & 10.01 & 7.81 & 0.35 & 0.12 \\
V338 Pup & $-$2.16 & $-$123.2 & 9.36  & 3.24 & 0.46 & 0.49 \\
V494 Sco & $-$1.01 & 210.9    & 8.54  & 6.73 & 0.21 & 0.12 \\
V690 Sco & $-$1.16 & 198.1    & 8.49  & 5.81 & 0.48 & 0.19 \\
  \hline
\end{tabular}
\end{center}
\end{table}

\begin{figure}
\begin{center}
   \epsfig{file=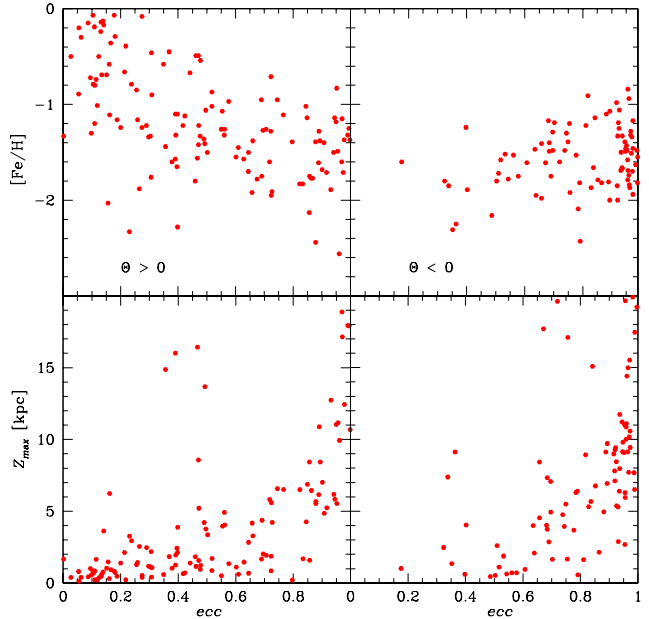,scale=0.45}
   \caption{Plots $ecc$ against [Fe/H] and   $Z_{\max}$.
    On the {\bf left} side  only \rrls\ with
     prograde orbits (${\Theta}>0 $)
    are plotted,
    on the {\bf right} side   only those with
     retrograde orbits (${\Theta}<0$).
    The eccentricities of the prograde \rrls\ spread
     over the hole range of $ecc$
    while the retrograde ones  show foremost higher eccentricities.
 }
   \label{ecc4.fig}
\end{center}
\end{figure}
A different way of looking at the data is given in Fig.\,\ref{ecc4.fig}.
There [Fe/H] and $Z_{\rm max}$ are plotted against $ecc$ for the prograde
and retrograde \rrls. Clearly, the continuous distribution indicated above
is present for the prograde set, while the retrograde stars rather are metal
poor and have orbits with predominantly larger eccentricity. In the
retrograde set there are only stars with [Fe/H]\,$<-0.8$ and $ecc>0.45$. We
conclude from this that the prograde set does contain a good fraction of
higher metallicity low-eccentricity orbit stars, in short younger \rrls,
while the retrograde set has only those with characteristics not like that
of a younger galactic disk.

We took a special look at the stars not reaching beyond $z_{\rm max}= 0.5$\,kpc
These stars are listed in Tab.\,\ref{05zmax.tab}.
Among these stars are two with a much smaller metallicity than the rest and
they turn out to have very eccentric orbits. These are \rrl\ itself, and 
V338 Pup whose orbit is retrograde. Thus 2 of 16 stars staying 
close to the plane turn out not to have disk-like orbits and thus 
belong to the halo population.

Chiba \& Beers (2000) published a list of likely metal-weak thick disk
stars including 3 \rrls\ using as criterium [Fe/H] $\leq -1$,
$ecc \leq 0.2$ and $183<\Theta<225$\,\kms.
In our sample we found 7 such \rrls: the three found by Chiba \& Beers (2000) 
plus TZ\,Aqr, RX\,Eri, V690\,Sco and AT\,Ser. 
We confirm that there are metal-weak stars with classical disk orbits,
which may be part of a metal-weak thick disk or in two cases (V494 Sco
and V690 Sco)  even thin disk star candidates having 
a $z_{\rm max}$$<$$0.5$\,kpc.  

\section{Halo distribution of RR Lyrae stars}
\label{otherstudies}

The spatial distribution of \rrls\ and other HB-like stars
has been investigated several times based on diverse samples.
In a pioneering study, Plaut (1965) derived a scale height of 
RR Lyr stars in the solar vicinity of $\simeq$2.3~kpc.
In his review Majewski (1993) reports scale heights for the 
``intermediate population II'', being $\simeq 1.5$\,kpc for \rrls.  
In almost all cases, the studies used photometric distances
perhaps together with radial velocity information.
Proper motions have thus far been 
included in limited cases only.

The vertical scale height derived in this paper from the orbit statistics
can be used for a comparison with results from other studies.

Kinman \ea (1994) used BHB stars
(with colours between the blue edge of the \rrl\ strip
and the intersection of the HB with the main sequence, so rather HBA stars)
to derive the structure of the halo from statistically complete samples.
They assume a (spherical) radial distribution of the form
\begin{equation}
 \rho = \rho_0 R_{\rm gal}^{-3.5}
\end{equation}
and find that the distribution of their HBA stars agrees with that.
We have fitted our $z$-distribution to the barometric formula
(Eq.\,\ref{barform}) and arrived at
a scale height of $\simeq5$\,kpc (see Fig.\,\ref{allergbest.fig}).
Transforming the run of densities of a radial distribution into an exponential
vertical distribution at the location of the Sun ($\varpi=8.5$\,kpc)
we find that the radial and the barometric relation (with $h=5$\,kpc)
give a run of densities $N(z)$ which, for $5<z<20$\,kpc,
deviate from each other by only 20\%
(the barometric formula being a bit steeper).
Thus our result from \rrl\ star orbits shows a spatial distribution
similar to the distribution of HBA stars from samples by Kinman \ea (1994).

The mean space density of \rrls\ in the solar neighbourhood has been derived
by Amrose \& Mckay (2001).
They arrive at different values for different solar distances but their
overall result is fairly summarised by $\rho=5\pm1$\, stars kpc$^{-3}$.
They also derive a vertical distribution and give fits
with the barometric formula allowing scale heights between
0.75 and 1.8\,kpc (using a single exponential fit).
However, from their Figure 5 it is clear that the \rrls\ show a
steeper distribution at low $z$ with $h\simeq1$\,kpc, similar to our result,
and a flatter one further out of $h>2$\,kpc.

Our data cannot be used to find an absolute mid-plane density. However,
one could use the ratio of
$N(0)_{\rm halo}/N(0)_{\rm disk \rrls}=0.16$  together with the
Amrose \& Mckay value of $\rho=5\pm1$ stars kpc$^{-3}$. Furthermore, the
midplane ratio may be combined with the scale heights derived for the disk
and halo components to get the ratio of all \rrls\ in these components. The
procedure is similar to that used by de Boer (2004) for sdB stars. We so
find that the Milky Way has about half as many halo \rrls\ than disk \rrls.
Since the midplane density ratio probably is still affected by systematic
effects and the scale heights are, among various studies, only just
consistent, the indicated total \rrl\ star ratio has to be regarded as a
crude value.

The data from the Sloan Digital Sky Survey (SDSS) revealed numerous HB stars
also at very large distances from the Milky Way disk
(see, e.g., Yanny \ea 2000, Sirko \ea 2004).
Vivas \ea (2001), using data from the {\sc quest} survey for \rrls,
find an over-density of \rrls\ at 50\,kpc from the Sun.
They surmise this over-density is part of the tidal stream
associated with the Sagittarius dwarf spheroidal galaxy.
From these new survey results we conclude that the halo is,
on a large scale, apparently rather inhomogeneous
so that \rrl\ distribution values will show to vary
in different directions and at different distances.

\section{Conclusions}
\label{conclusions}
\rrls\ exist in the Milky Way as part of the disk and part of the halo. 
We separated the halo component from disk candidates using the shape of 
their orbits. Of 217 \rrls, 154 or 71~\% are member of the halo population. 
Therein we find a retrograde rotating halo component
of 87 \rrls\ or more than half of our halo group. 
The (vertical) spatial distribution found with our orbit statistics method is,
out to $z~=20$~kpc,
similar to that found in other studies using classical methods.
We show that there are \rrls\ staying in the disk
 but which are member of the halo population.  \\ 

\begin{acknowledgements}
The orbit programme was put at the disposal of the Bonn group by 
Michael Odenkirchen. 
We thank Martin Altmann for providing several plotting programs as well as 
constructive criticism and Torsten Kaempf and Michael Geffert 
for helpful discussions and the referee, Tim Beers for suggestions 
that improved the paper. 
\end{acknowledgements}

\begin{table}
\setlength{\tabcolsep}{1.35mm}  
\caption[]{Electronic Table;  Positions, velocities (given in the Galactic euclidic system $XYZUVW$ and $\Theta, \Phi$)
and orbital data ($ R_{\rm a}, R_{\rm p}, z_{\rm max}$, normalised $z$-extent, $nze$ and  eccentricy $ecc$) of the stars of our sample.
This table is only available in electronic form at the CDS via anonymous ftp to cdsarc.u-strasbg.fr (130.79.128.5)
or via http://cdsweb.u-strasbg.fr/cgi-bin/qcat?J/A+A/.
}
\label{tablecds}

\begin{tabular}{lrrrrrrrrrrrrr}
\hline
 \hline
Name &  {\it X}&  {\it Y} &  {\it Z} &  {\it U}  &  {\it V} &  {\it W} & {$\Phi$}&  {$\Theta$} &   $R_{\rm a}$ &  $R_{\rm p}$ &  $z_{\rm max}$ &  $nze$ &  $ecc$ \\
 &  \multicolumn{3}{l}{\ \ [kpc]} &  \multicolumn{5}{l}{\ \ [km\,s$^{-1}$]}   &  \multicolumn{3}{l}{\ \ [kpc]} &$-$  & $-$ \\
 \hline

\end{tabular}
Online-Data are available under http://www.astro.uni-bonn.de/$\sim$gmaintz

\end{table}

\end{document}